\documentclass{webofc}
\usepackage[varg]{txfonts} 
\begin{document}

\title{Portable Resistive Plate Chambers for Muography in confined environments}
% \subtitle{1st draft}

\author{\firstname{R.M.I.D} \lastname{Gamage}\inst{1}\fnsep\thanks{\email{ishan.ran@uclouvain.be}}
\and \firstname{Samip} \lastname{Basnet}\inst{1}
\and \firstname{Eduardo} \lastname{Cortina Gil}\inst{1}
\and \firstname{Andrea} \lastname{Giammanco}\inst{1}
\and \firstname{Pavel} \lastname{Demin}\inst{1}
\and \firstname{Marwa} \lastname{Moussawi}\inst{1}
\and \firstname{Amrutha} \lastname{Samalan}\inst{2}
\and \firstname{Michael} \lastname{Tytgat}\inst{2}
\and \firstname{Raveendrababu} \lastname{Karnam}\inst{1,3} 
\and \firstname{Ayman} \lastname{Youssef}\inst{4}
}

\institute{Centre for Cosmology, Particle Physics and Phenomenology (CP3), Université Catholique de Louvain, Louvain-la-Neuve, Belgium 
\and
            Department of Physics and Astronomy, Ghent University, Ghent, Belgium
\and
            Centre for Medical and Radiation Physics (CMRP), National Institute of Science Education and Research (NISER), Bhubaneswar, India
\and
           Multi-Disciplinary Physics Laboratory, Optics and Fiber Optics Group, Faculty of Sciences, Lebanese University, Hadath, Lebanon
          }

\abstract{%
 
Muography (or muon radiography) is an imaging technique that relies on the use of cosmogenic muons as a free and safe radiation source. It can be applied in various fields such as archaeology, civil engineering, geology, nuclear reactor monitoring, nuclear waste characterization, underground surveys, etc. In such applications, sometimes deploying muon detectors is challenging due to logistics, e.g. in a narrow underground tunnel or mine. 
Therefore, we are developing muon detectors whose design goals include portability, robustness, autonomy, versatility, and safety. Our portable muon detectors (or ``muoscopes'') are based on Resistive Plate Chambers (RPC), planar detectors that use ionization in a thin gas gap to detect cosmic muons. 
Prototype RPCs of active area $16 \times 16~cm^2$ and $28 \times 28~cm^2$ were built in our laboratories at Louvain-la-Neuve (UCLouvain) and Ghent (UGent) to test and compare various design options. Benefiting from the experience gained in building and operating these prototypes, we are proceeding towards the development of improved prototypes with more advanced technical layout and readiness. In this paper we provide the status of our performance studies, including the cross-validation of the two types of prototypes in a joint data taking, and an outline of the direction ahead.

% Keywords : Particle Detectors, Resistive plate chambers, Imaging techniques, Muography
}
\maketitle

%  \tableofcontents
%

\section{Introduction}
\label{intro}
Muography, i.e. muon-based radiography, is an imaging technique based on the absorption or scattering of atmospheric muons. These muons are produced when primary cosmic rays interact with the upper Earth's atmosphere. Muons are insensitive to strong nuclear interactions, and thanks to their mass roughly 200 times larger than the electrons, they have a low rate of energy loss by ionization and other electromagnetic processes. At the energies that are typical of atmospheric muons at sea level, muons have a high penetration power, which makes them an appropriate source for the radiography of large volumes.\\
Usage of high-energy cosmic muons for imaging dates back to the mid-20th century~\cite{George1955}, when the attenuation of the muon flux (measured with a movable Geiger counter telescope) was used to probe the overburden of a tunnel. Since then, muography found various applications in different fields ~\cite{Bonechi2019ckl} such as volcanology, archaeology, civil engineering, and several applications in the nuclear and security sectors. To study volcanoes and pyramids, given their size, muon detectors must have large cross-sectional area. In other cases however, which motivate the specific design choices of the project reported in this paper, the optimal viewpoint may be located in narrow and confined environments such as tunnels or underground chambers, which set tight constraints on the size of the detectors.
Use cases for portable detectors can also be found in cultural heritage preservation, e.g. statues hosted in crowded museal spaces and which are too heavy or too fragile to be transported to a laboratory.
A few teams are developing compact muography detectors for usage underground. In some cases, very specialized detectors with cylindrical symmetry are developed that fit inside boreholes, as first proposed in \cite{Malmqvist1979}. 
More traditional geometries can be used when a pre-existing tunnel can be exploited. 
Examples of portable detector projects include MIMA~\cite{Baccani:2018nrn}, based on scintillating bars coupled with silicon photomultipliers, a project in Kyushu University~\cite{Kyushu2018} based on thin scintillating fibers, and MUST2~\cite{Roche_2020}, a time-projection chamber based on micromegas detectors. 
For our project, we chose to explore the feasibility of small-area Resistive Plate Chambers (RPCs) for muography, motivated by their good trade-off between cost and simplicity of construction (making them promising for mass production) versus performance metrics such as spatial and time resolution, efficiency and fake rate.

An RPC is a planar detector with parallel plate geometry that uses a mixture of gases as the active detection medium~\cite{Santonico-Cardarelli_1981,AbbresciaPeskovFonte_2018}. It consists of two highly resistive plates which are placed parallel to each other, with semi-conducting coating on the exterior sides of the plates. The volume between the two high resistive plates is filled with an appropriate gas mixture, and high voltage (HV) is supplied to the two resistive layers. RPCs have already been used in a few muography applications~\cite{Menedeu:2016rrr,Baesso_2013,TUMUTY_pan_2019}; their challenges in this context have been critically discussed in \cite{MuographyBook}. 

Our overarching goals include portability, compactness, low weight, low cost, autonomy and versatility. 
This paper reports on the prototypes built at UCLouvain and UGent in Belgium, and on the results of preliminary tests to assess their performance. 
Previous steps in our project have been reported in \cite{Wuyckens2018,Basnet2020,Gamage_2022,Moussawi2021,Basnet:2022cds}. Our prototype chambers have active areas of either $16 \times 16~cm^2$ or $28 \times 28~cm^2$ and differ in a few design choices, 
as elaborated later.
Using the approximate formulas in \cite{lechmann2021muon}, under the assumption of ideal detector performance, for a depth of 50~$m$ of rock the active area of the smallest of these two prototypes would potentially achieve a lateral spatial precision of 5\% to discern density differences of more than 15\% after 3 days of data taking, and a lateral spatial precision of 1\% after 100 days. 

%As RPCs are gas detectors, leaks are a concern in confined spaces; moreover, we need to minimize refills. That led to the development of sealed casings, which in turn imposed attention to potential gas contamination from outgassing of materials. 

This paper is organized as follows. 
In Section~\ref{sec:RPC-in-Confined-Env}, we motivate our design goals.
Our current RPC prototypes are described in Section~\ref{sec:Prototypes}, while Section~\ref{sec:Results} presents the results of some tests. 
Finally, Section~\ref{sec:future} introduces work ongoing towards more mature prototypes.

\section{RPC Detectors for Confined Environments}
\label{sec:RPC-in-Confined-Env}

%\subsection{Design Goal}
%\label{Design}
We are developing a small-scale RPC-based Muon Detector which we named Muoscope. The design goal of the detector includes portability; safety; robustness, i.e. ability to work under different conditions; and autonomy, which means the detector should be able to operate for a long period without the active involvement of a human operator.
% \begin{figure}
% % Use the relevant command for your figure-insertion program
% % to insert the figure file.
% \centering
% \includegraphics[width=8cm]{images/Mining_F.png}
% \centering
% \caption{Illustration of Muography for a mining survey}
% \label{fig-1}       % Give a unique label
% \end{figure}
% 

%\subsection{Muography in Confined Environments}
%\label{Confined-Env}
Muography in confined environments has certain specific challenges such as limited space for installing the detector setup (which motivates our focus on small-area RPCs), and potentially no power available. %We are aiming to address these limitations with our detector setup. 
Portability imposes restrictions also on the weight of the whole setup, including all passive material (RPC casings) and any ancillary equipment. 
To achieve autonomy we aim at powering the whole set-up (including the Data Acquisition system) with batteries, therefore we aim at a very low-power readout and control electronics, to be also able to adapt to changing environmental conditions. 
In general for gaseous detectors, such as RPCs, human safety is an important concern, as most commonly used gas mixtures are toxic and flammable; moreover, a massive leak from a gas bottle used to flush the detector~\cite{procureur2020we} may lead to hypoxia. This motivates sealed gas chambers~\cite{Assis:2020mvp}, which also reduce the need for gas refills and therefore contribute to autonomy. 

\section{Current Prototype Detectors}
\label{sec:Prototypes}

We are developing two sets of RPCs with slightly different characteristics, cross-validating them in view of future co-developed detectors for muography applications. RPCs are gaseous detectors, where a suitable gas mixture acting as the detection medium fills a thin gap between two high-resistivity plates. The exterior sides of the plates are coated with semi-resistive paint (carbon-based in our case~\cite{Basnet2020}), whose purpose is to distribute the HV evenly throughout the resistive surfaces (glass plates in our prototypes) in order to get a homogeneous electric field through the gap. When a charged particle passes through, it ionizes the gas molecules along its path, releasing free electrons that accelerate toward the electrode, collide with gas molecules and produce an avalanche. The motion of these charges induces an electric signal in the metallic readout strips, which are read individually. 
Our prototypes all use the same gas mixture: R134a Freon (95.2\%), isobutane (4.5\%) and SF$_{6}$ (0.3\%).

\begin{figure}[htbp]
\centering
(a)
\includegraphics[width=5cm]{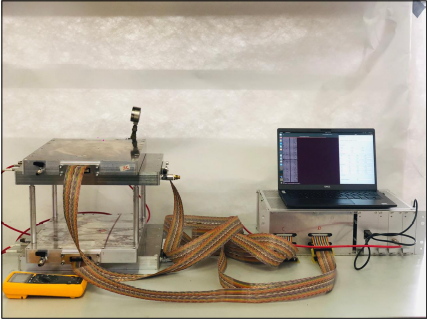}
(b)
\includegraphics[width=5cm]{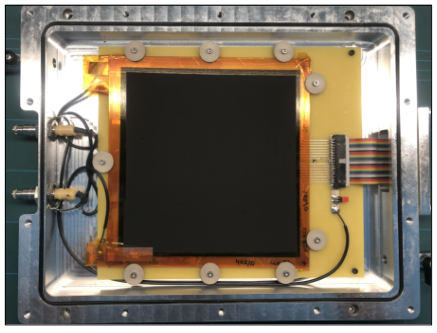}
\caption{\label{fig:detector1.L} (a) Muoscope Prototype 1.L.     (b) One of the RPCs of Prototype 1.L inside its casing.}
\end{figure}

%\subsection{Prototype 1.L}
%\label{1.L}
\paragraph{Prototype 1.L}
The 1.L prototype consists of four identical RPCs, each hosted in a gas-tight aluminium box, and a Data Acquisition (DAQ) system integrated with a high-voltage power supply. As shown in Figure \ref{fig:detector1.L}(a), the first and third RPCs are oriented orthogonal to the second and fourth, in order to provide bi-dimensional information (X and Y orientations), with spacers used to set a 14.8~$cm$ distance between second and third. 
Offline, we reconstruct a muon trajectory by using the X and Y information at the top and at the bottom of this setup. 
Each detector layer (casing + RPC) weighs 6.5~$kg$.

Each RPC consists of two glass plates with a thickness of 1.1~$mm$ and a gas gap of 1~$mm$. 
A uniform distance between the glass plates was obtained by placing nine-round edge spacers made of polyether ether ketone (PEEK). 
A uniform surface resistivity of $4~M\Omega$/$\square$ was obtained on the glass plates by spreading the semi-resistive paint using serigraphy~\cite{Basnet2020}, and the evolution of the resistivity with time has been monitored for more than two years~\cite{Gamage_2022}. 
The readout board of the RPC is made with sixteen copper strips. Each strip is 9~$mm$ wide and separated by a 1~$mm$ gap, thus yielding a pitch of 1~$cm$. %Each RPC is hosted in an air-tight aluminum box. The rate of gas leakage in vacuum conditions was measured using helium to be 10$^{-9}$ mbar~l~s$^{-1}$~\cite{Wuyckens2018}. 
 %The current gas mixture consists of R134a Freon (95.2\%), isobutane (4.5\%) and SF$_{6}$ (0.3\%), kept at a pressure slightly above (by $\sim$0.1 atm) the atmospheric one.
The DAQ electronics consist of two front-end boards (FEBs) borrowed from the RPC system of the CMS experiment~\cite{FEB1,FEB2}. Each FEB can handle 32 analog inputs channels, each consisting of an amplifier with a charge sensitivity of $2~mV/fC$, a discriminator, a monostable and a LVDS driver. The LVDS outputs of all the FEBs are connected to a System-on-Chip module, which is installed on a carrier board with a wireless connection, to ensure autonomy.

%\subsection{Prototype 1.G}
%\label{1.G}

\paragraph{Prototype 1.G}
The Prototype 1.G consists for the moment of a single chamber, although more will be produced. It has an active area of $28 \times 28~cm^2$ and the glass resistive plates are of thickness 1.3~$mm$. A mixture of a conductive and resistive paint was used for obtaining the resistive layer on the glass plates and then an in-house spray gun based technique was used for coating. A uniform resistive layer in terms of thickness as well as resistivity ($450~k\Omega$/$\square$) was achieved, which is up to the quality level of industrially developed resistive plates. The gas gap of thickness 1~$mm$ is obtained by placing three fish wires between the glass plates separated in equal distance. The PCB board for the signal readout consists of 16 copper strips of width 15~$mm$ and separated by a distance of 1.6~$mm$, thus yielding a pitch of 16.6~$mm$. The whole setup is placed in a honeycomb-based gas-tight box. The box consists of a HV connector, gas inlet, gas outlet and a slot to connect the flat cables to the readout board. The gas is allowed to flow inside the whole system and at the same time gas at slow flow rate is supplied to the chamber to refresh the gas inside. The CAEN standard HV module is used to power the chamber and the signal processing is done using a CAEN based 32 channel ADC. 
\begin{figure}[htbp]
\centering
(a)
\includegraphics[width=5cm]{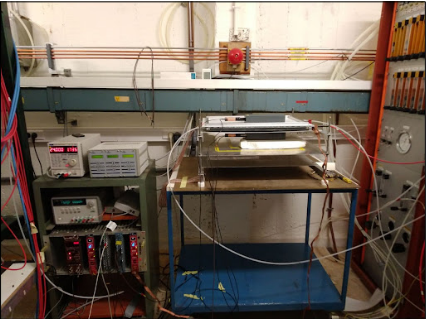}
(b)
\includegraphics[width=5cm]{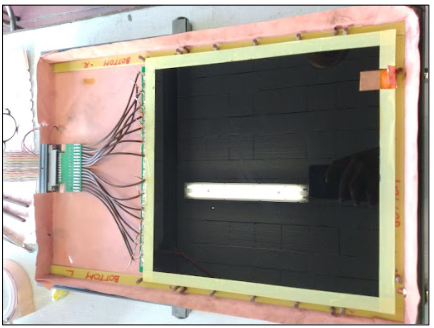}
\caption{\label{fig:detector1.G} (a) RPC Prototype 1.G    (b) RPC detector inside its casing.}
\end{figure}

The main differences between prototypes 1.L and 1.G are summarized in Table~\ref{tab:comparison}.

\begin{table}[ht]
\centering
\begin{tabular}{|c|c|c|}
\hline
{\bf Property} & {\bf Prototype 1.G} & {\bf Prototype 1.L} \\ \hline
\hline
Active area & $28 \times 28~cm^2$ &$16 \times 16~cm^2$  \\ \hline
Gas Flow & Continuous & Sealed  \\ \hline
%Gas Mixture & 95.2\% Freon, 0.3\% SF_{6} & 95.2\% Freon, 0.3\% SF$_{6}$ \\
%& 4.5\% isobutane & 4.5\% isobutane \\
%\hline
Glass thickness & $1.3~cm$ & 1.1~$cm$ \\ \hline
Strip Width & 15~$mm$ & 9~$mm$ \\ \hline
Strip Pitch & 16.6~$mm$ & 10~$mm$ \\ \hline
Semi-resistive & Hand-sprayer & Serigraphy \\ 
coating &  ($\sim$ $450~K\Omega$/$\square$) &  ( $\sim$ $4~M\Omega$/$\square$)\\ \hline
DAQ & NIM + CAEN integrated & Custom made\\ \hline
Portability & Not yet & Portable\\ \hline
\end{tabular}
%\vskip 0.1cm
\label{tab:comparison}
\caption{Main differences between the two RPC prototypes developed so far.}
\end{table}

\section{Results}
\label{sec:Results}

In this section, we are describing the preliminary results from our prototypes. The individual results are not in the same format because each prototype has its own DAQ and the ways they record data are different. But in the joint data taking of the two detectors, we are using the DAQ of prototype 1.L and so we could validate the two sets of data.

\paragraph{Prototype 1.L}
%\label{Results_1.L}

\begin{figure}[htb]
\centering
(a)
\includegraphics[width=7cm]{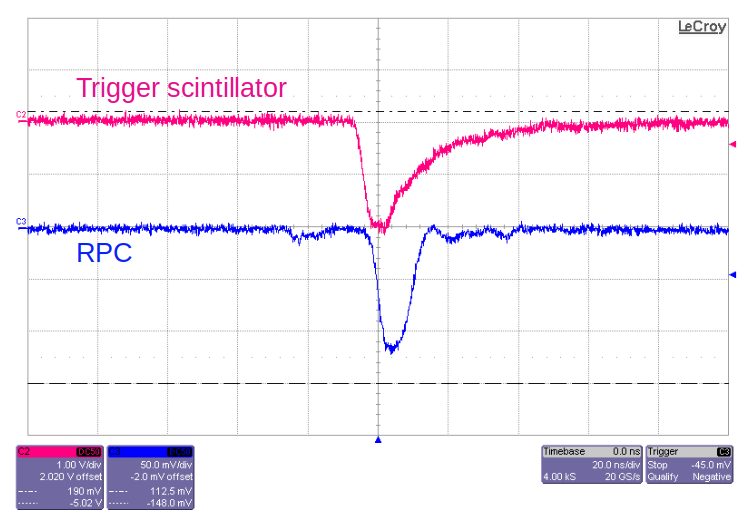}
(b)
\includegraphics[width=4.5cm]{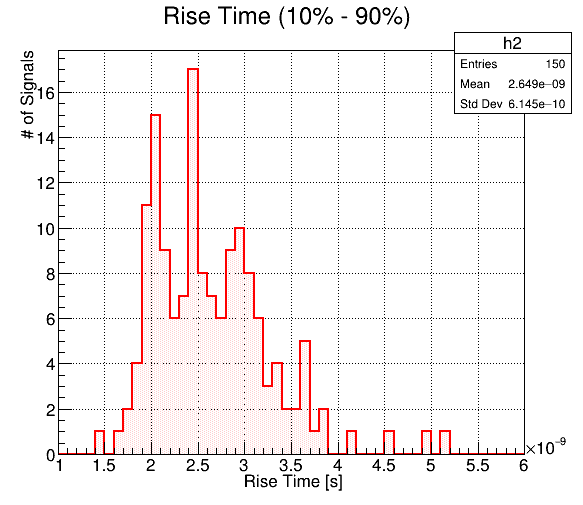}
\caption{\label{fig:RPCSignal} (a) Oscillogram of the RPC signal (blue) and of the signal from one of the scintillators used as external trigger (red). (b) Rise time distribution of the RPC signal measured using the oscilloscope.}
\end{figure}

We intend to operate our RPCs in avalanche mode, to get fast signals with a rise time of a few nano seconds. To verify the operating mode of the 1.L RPCs, we checked the signal from single strips, triggering on signals from external scintillator detectors, see Fig.~\ref{fig:RPCSignal}(a). Using the oscilloscope, we measure on average a signal amplitude of $\sim$~-160~$mV$ and a rise time of $\sim$~2.65~$ns$. Figure \ref{fig:RPCSignal}(b) shows the distribution of rise times. We conclude that an operating voltage of 7~$kV$ is appropriate under the current conditions, but further studies may lead to fine adjustment of the working point.

\paragraph{Prototype 1.G}
%\label{Results_1.G}

\begin{figure}[htb]
\centering
(a)
\includegraphics[width=5cm]{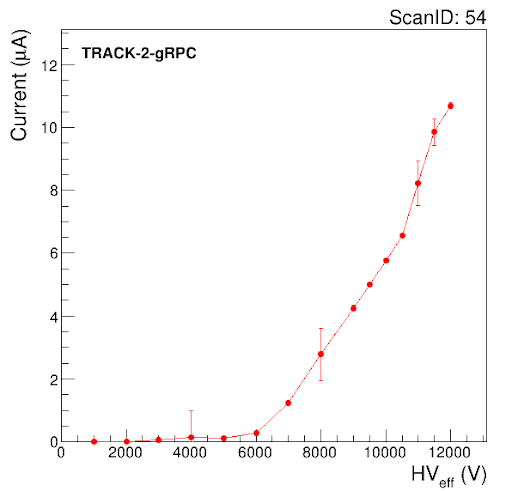}
(b)
\includegraphics[width=5cm]{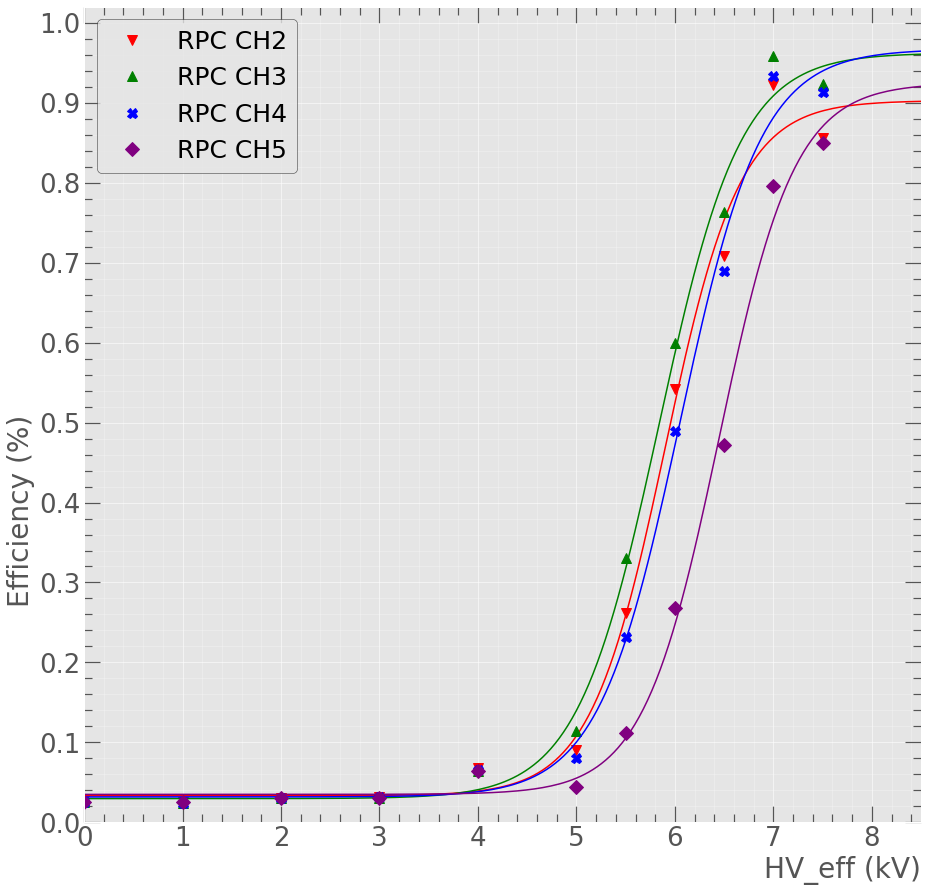}
\caption{\label{fig:UGhent} (a) Ohmic behavior of 1.G. (b) Rise time of the signal measured using the oscilloscope.}
\end{figure}

To determine the operating HV of the avalanche mode in Prototype 1.G, we used the Ohmic behavior of the detector. The current drawn by the detector was measured as function of HV, see Fig.~\ref{fig:UGhent}(a). In avalanche mode the current drawn by detector behaves linearly with respect to HV. This determined a working point between 6~$kV$ and 8~$kV$. This range of HV is large, so it needed a fine adjustment. The efficiency of the each strip of the RPC is measured for different HV values with an external trigger scintillator. The efficiency of each strip is calculated using two external trigger scintillators. Efficiencies of 4 different strips are shown in Figure \ref{fig:UGhent}(b). We chose 7~$kV$ as working point, yielding more than 80\% efficiency.

\paragraph{Joint data taking of 1.L and 1.G}

\begin{figure}[htpb]
\centering
(a)
\includegraphics[width=5cm]{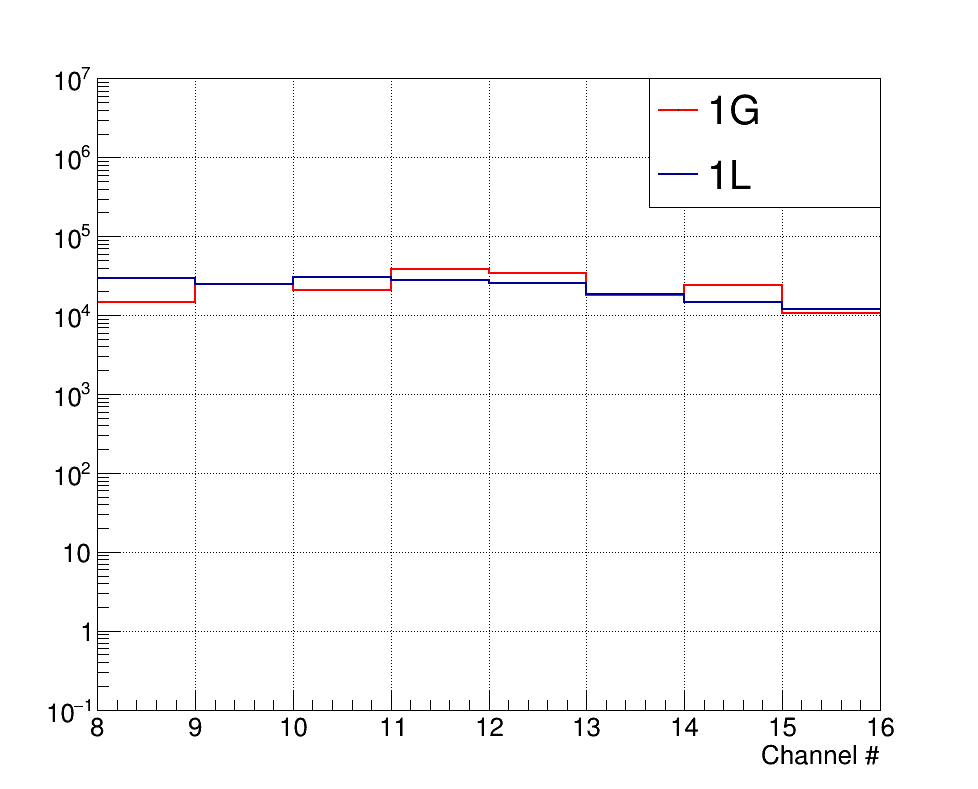}
(b)
\includegraphics[width=5cm]{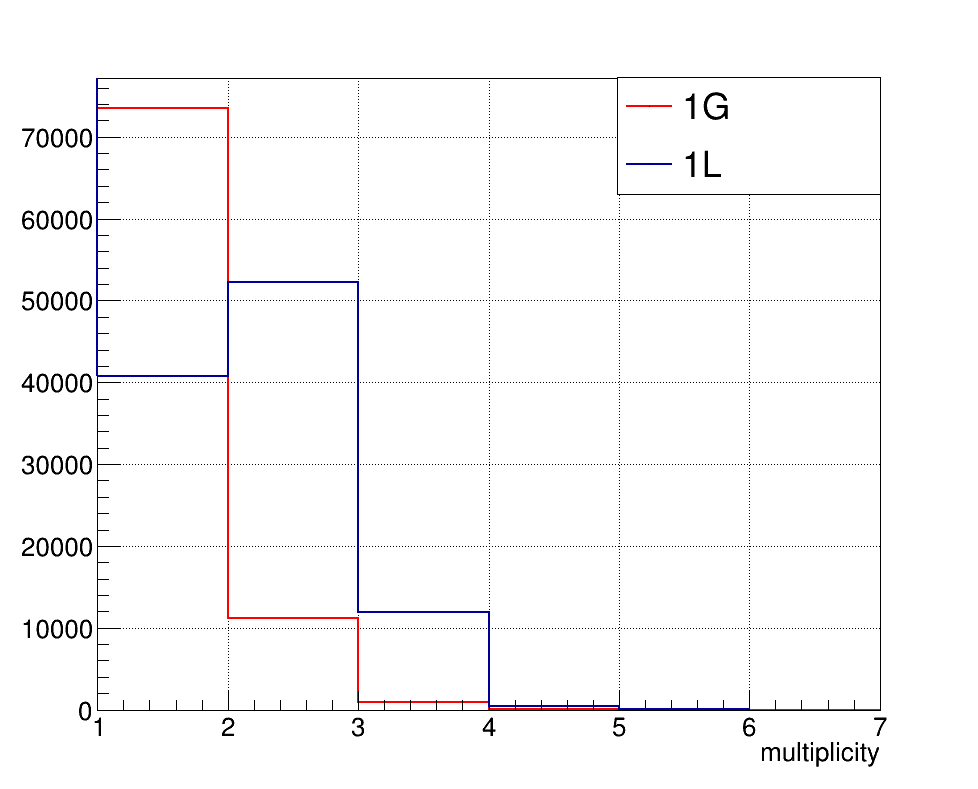}
\caption{\label{fig:JointData} (a) Occupancy: number of times each strip fired throughout
the data-taking run. (b) Multiplicity: total number of strips fired per event.}
\end{figure}

We made a joint data taking of two prototypes connected to the DAQ of Prototype 1.L in order to cross-validate their performances. Working points were set to our first estimated values which are HV $=$ 7~$kV$ and a threshold discriminator value of 90 DAQ units. In Fig.~\ref{fig:JointData} the occupancy (i.e. number of times each strip fired throughout the run) and the multiplicity (i.e. total number of strips fired per event) are shown. Although both detectors have 16 strips, the plots only shows the result from the last 8 strips, as the first 4-5 strips in the Prototype 1.G were affected by hardware related issues that were fixed only after this data collection.

In Fig.~\ref{fig:JointData}, both distributions are normalized to the 1.L active area. The occupancy distributions of both prototypes are in general agreement. So we can conclude the two detectors behave in the same way, which cross-validates each. However, we observe a difference in the  multiplicity distribution, which is higher in 1.G. This could be explained with the difference in the resistivity of the glass plates, which is smaller for 1.G than 1.L. 

\section{Towards Muoscope 2.L}
\label{sec:future}

Taking stock of the experience in building and operation of a small-area Muoscope that we gained with Prototype 1.L, we are working in parallel to lay the ground to a future Prototype 2.L. A few options are being investigated in parallel, as discussed in this section.

\begin{figure}[htpb]
\centering
\includegraphics[width=10cm]{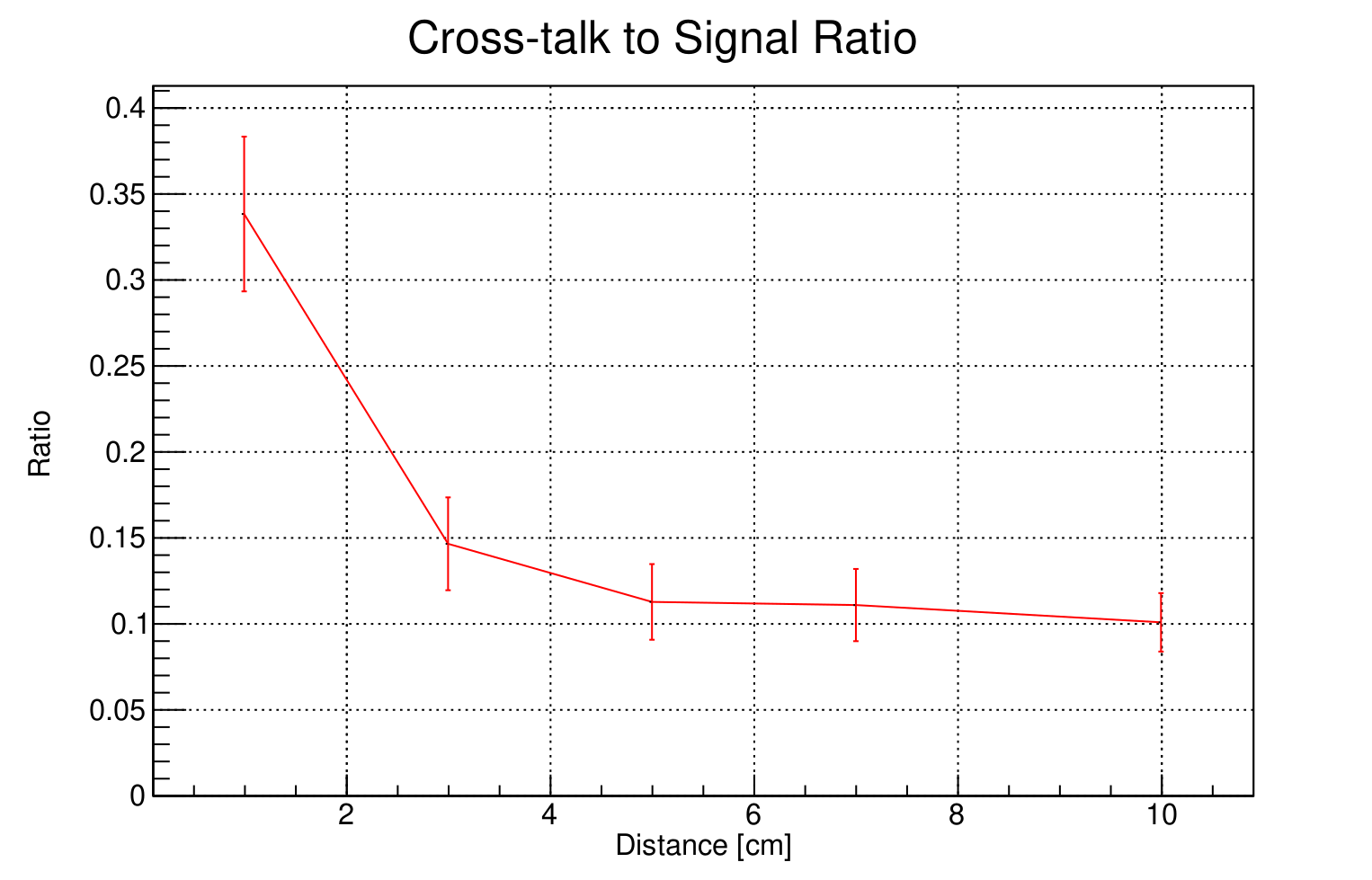}
\caption{\label{fig:crosstalk} Cross-talk to signal ratio with respect to the distance from the respective readout strip.}
\end{figure}

\begin{figure}[htpb]
\centering
(a)
\includegraphics[width=5cm]{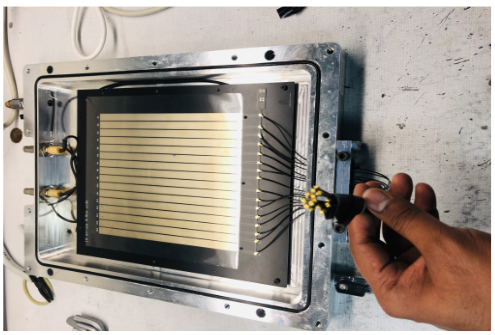}
(b)
\includegraphics[width=4cm]{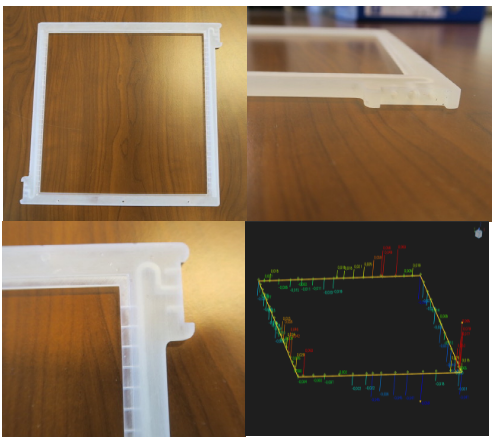}
\caption{\label{fig:2.L} (a) Readout board with coaxial cables connected to strips. (b) 3D-printed frame.}
\end{figure}

\paragraph{Readout board}
The current readout board consists of thin copper tracks, $\rm \mathcal{O}(1mm)$ wide, connected to the readout strips and 16 flat cables to deliver signal to the DAQ. 
A mismatch of the impedance of the thin tracks and flat cables causes large reflection of the analog RPC signal. Moreover, we observe large cross talk between strips, as seen in Fig.~\ref{fig:crosstalk}; in particular, neighboring strips have a cross talk of 35\%. 
In the new readout board, copper tracks and flat cables are replaced by coaxial cables mounted directly to the readout strips, as shown in Fig.~\ref{fig:2.L}(a). 
To address the above mentioned issues and reduce noise, we will use high-end materials and also improve the grounding.

\paragraph{3D-printed frames}

In Prototype 1.L, the distance between the two glass plates is kept uniform by 9 round spacers at the edges of the glass plates, see Fig.~\ref{fig:detector1.L}(b). 
For the next prototype, we are considering the option of 3D-printing a rigid frame of a light material (carbon fibre or plastic), to which the glass plates would be glued, ensuring a uniform gap.
It will be air-tight and the gas needed for the RPC would be much less, as only the volume between the glass plates would be filled. 
This 3D-printed frame would also replace the aluminum casing of 1.L, with an inlet and an outlet for the gas, designed to ensure a uniform gas distribution throughout the gap. 
Being of a much lighter material than aluminum, it would improve the portability of the whole setup. 
Our first test frame, in plastic, is shown in Fig.~\ref{fig:2.L}(b). 

Although all these characteristics are very appealing, a thorough dedicated study will be needed to quantify the possible contamination of the operating gas due to outgassing from the casing itself~\cite{procureur2020we}, which is a major concern given the requirement of no constant flushing.

\paragraph{From strips to pixels}
We are considering an ambitious option for future prototypes: moving from strips to pixels. 
The possibility to extract 2D information from a single RPC instead of two orthogonal ones, and therefore halving the number of detection layers, is appealing because 
the overall efficiency of the muoscope would increase ($\epsilon^{N/2}$ > $\epsilon^N$, where $\epsilon$ is the single-RPC efficiency and $N$ is the number of RPCs) while also reducing the total weight. 

This option, however, has a major drawback. With pixels, both cost and power consumption of the electronics scale quadratically with the number of channels. 
Already with our current resolution, two pixel layers (2$\times$16$^{2}$=512 readout channels) would be an order of magnitude more pricey in both cost and power consumption compared to the current strip muoscope (4$\times$16=64 channels). 
When improving the spatial resolution of future prototypes, the difference between the two options will become even larger. 
Cost and power consumption are important considerations in this project, in particular the latter (of the order of $mW$/channel) affects autonomy by the possibility, or not, to operate with batteries or solar panels.

An appropriate solution is reading out several units with just one electronic channel, i.e. multiplexing. We are currently developing a 2D-specific multiplexing / de-multiplexing approach, the first steps of which have been reported in \cite{Basnet:2022cds}.

\section{Conclusion}

To summarize this paper, this is the first time we are presenting the result from joint data taking of Prototypes 1.L and 1.G, which were built with slightly different characteristics, to gain construction experience and compare different options. 
The current detector performances encourage us to proceed and to start laying the ground to more ambitious prototypes. 

\section*{Acknowledgements}

This work was partially supported by 
%the EU Horizon 2020 Research and Innovation Programme under the Marie Sklodowska-Curie Grant Agreement No. 822185, and by
the Fonds de la Recherche Scientifique - FNRS under Grants No. T.0099.19 and J.0070.21. Samip Basnet and Raveendra Babu Karnam would like to acknowledge additional research grants from FNRS FRIA and UCLouvain Special Research Funds, respectively. 
We gratefully acknowledge the work performed by Sophie Wuyckens between 2017 and 2019, which was crucial for the kick-off of this R\&D project.

\bibliography{refs}

\begin{thebibliography}{22}

\bibitem{George1955}
E.P. George, Commonwealth Engineer \textbf{July 1}, 455 (1955)

\bibitem{Bonechi2019ckl}
L.~Bonechi, R.~D’Alessandro, A.~Giammanco, Reviews in Physics \textbf{5},
  100038 (2020)

\bibitem{Malmqvist1979}
L.~Malmqvist, G.~J\"{o}nsson, K.~Kristiansson, L.~Jacobsson, Geophys.
  \textbf{44(9)}, 1549 (1979)

\bibitem{Baccani:2018nrn}
G.~Baccani et~al., JINST \textbf{13}, P11001 (2018), \texttt{1806.11398}

\bibitem{Kyushu2018}
K.~Chaiwongkhot et~al., IEEE Trans. Nucl. Sci. \textbf{65}, 2316 (2018)

\bibitem{Roche_2020}
I.~{L{\'{a}}zaro Roche}, J.B. Decitre, S.~Gaffet, Journal of Physics:
  Conference Series \textbf{1498}, 012048 (2020)

\bibitem{Santonico-Cardarelli_1981}
R.~Santonico, R.~Cardarelli, Nuclear Instruments and Methods \textbf{187}, 377
  (1981)

\bibitem{AbbresciaPeskovFonte_2018}
M.~Abbrescia, P.~Fonte, V.~Peskov, \emph{Resistive Gaseous Detectors - Designs,
  Performance, and Perspectives} (Wiley-VCH, Weinheim, 2018)

\bibitem{Menedeu:2016rrr}
E.~{Le Menedeu} (TOMUVOL), JINST \textbf{11}, C06009 (2016),
  \texttt{1605.09218}

\bibitem{Baesso_2013}
P.~Baesso, D.~Cussans, C.~Thomay, J.J. Velthuis, J.~Burns, C.~Steer,
  S.~Quillin, Journal of Instrumentation \textbf{8}, P08006 (2013)

\bibitem{TUMUTY_pan_2019}
X.Y. Pan, Y.F. Zheng, Z.~Zeng, X.W. Wang, J.P. Cheng, Nuclear Science and
  Techniques \textbf{30}, 120 (2019)

\bibitem{MuographyBook}
A.~Giammanco, E.~Cortina~Gil, S.~Andringa, M.~Tytgat, in \emph{{Muography:
  Exploring Earth's Subsurface with Elementary Particles}}, edited by L.~Olah,
  H.~Tanaka, D.~Varga (AGU - Wiley, 2021), Geophysical Monograph Series (ISSN:
  0065-8448), chap.~18

\bibitem{Wuyckens2018}
S.~Wuyckens, A.~Giammanco, P.~Demin, E.~{Cortina Gil}, Phil. Trans. R. Soc. A
  \textbf{377}, 0139 (2018), \texttt{1806.06602}

\bibitem{Basnet2020}
S.~Basnet, E.~{Cortina Gil}, P.~Demin, R.~Gamage, A.~Giammanco, M.~Moussawi,
  M.~Tytgat, S.~Wuyckens, JINST \textbf{15}, C10032 (2020), \texttt{2005.09589}

\bibitem{Gamage_2022}
R.~Gamage, S.~Basnet, E.~{Cortina Gil}, P.~Demin, A.~Giammanco, R.~Karnam,
  M.~Moussawi, M.~Tytgat, Journal of Instrumentation \textbf{17}, C01051 (2022)

\bibitem{Moussawi2021}
M.~Moussawi, S.~Basnet, E.~{Cortina Gil}, P.~Demin, R.~Gamage, A.~Giammanco,
  R.~Karnam, A.~Samalan, M.~Tytgat, \emph{{A portable muon telescope for
  exploration geophysics in confined environments}}, in \emph{{First
  International Meeting for Applied Geoscience \& Energy, 26 September - 1
  October 2021, Denver (USA)}} (2021), SEG Technical Program Expanded
  Abstracts, First International Meeting for Applied Geoscience \& Energy
  Expanded Abstracts, 3034-3038

\bibitem{Basnet:2022cds}
S.~Basnet, E.~{Cortina Gil}, P.~Demin, R.~Gamage, A.~Giammanco, R.~Karnam,
  M.~Moussawi, A.~Samalan, M.~Tytgat, \emph{{Towards a portable high-resolution
  muon detector based on Resistive Plate Chambers}}, in \emph{{International
  Workshop on Cosmic-Ray Muography}} (2022), \texttt{2202.01084}

\bibitem{lechmann2021muon}
A.~Lechmann, D.~Mair, A.~Ariga, T.~Ariga, A.~Ereditato, R.~Nishiyama,
  C.~Pistillo, P.~Scampoli, F.~Schlunegger, M.~Vladymyrov, Earth-science
  reviews \textbf{222}, 103842 (2021)

\bibitem{procureur2020we}
S.~Procureur, D.~Atti{\'e}, S.~Bouteille, D.~Calvet, X.~Coppolani, B.~Gallois,
  H.~Gomez, M.~Kebbiri, E.~Le~Courric, P.~Magnier et~al., Nuclear Instruments
  and Methods \textbf{955}, 163290 (2020)

\bibitem{Assis:2020mvp}
L.~Lopes et~al., JINST \textbf{15}, C11009 (2020), \texttt{2006.08291}

\bibitem{FEB1}
M.~Abbrescia, A.~Colaleo, G.~Iaselli, F.~Loddo, M.~Maggi, B.~Marangelli,
  S.~Natali, S.~Nuzzo, G.~Pugliese, A.~Ranieri et~al., Nuclear Instruments and
  Methods \textbf{456}, 143 (2000)

\bibitem{FEB2}
C.~Binetti, R.~Liuzzi, F.~Loddo, B.~Marangelli, A.~Ranieri, CMS Note-1999/047
  (1999)

\end{thebibliography}

\end{document}